\documentclass[]{spie}  %>>> use for US letter paper
\usepackage{graphicx,amsmath,amssymb,latexsym} 
\newcommand{\farcs}{\mbox{\ensuremath{.\!\!^{\prime\prime}}}}

\def\apj{ApJ}%
\def\apjl{ApJ}%
\def\ao{Appl.~Opt.}%
\def\aap{A\&A}%
\def\nat{Nature}%

\title{L'-band AGPM vector vortex coronagraph's first light on LBTI/LMIRCam}

\author{D.~Defr\`ere\supit{a}, O.~Absil\supit{b}, P.~Hinz\supit{a}, J.~Kuhn\supit{c}, D.~Mawet\supit{d}, B.~Mennesson\supit{c}, A.~Skemer\supit{a}, J.~Kent Wallace\supit{c}, V.~Bailey\supit{a},  E.~Downey\supit{a},  C.~Delacroix\supit{b},  O.~Durney\supit{a},  P.~Forsberg\supit{e},  C.~Gomez\supit{b},  S.~Habraken\supit{b},  W.F.~Hoffmann\supit{a}, M.~Karlsson\supit{e},  M.~Kenworthy\supit{f},  J.~Leisenring\supit{a},  M.~Montoya\supit{a},  L.~Pueyo\supit{g},  M.~Skrutskie\supit{h}, and J.~Surdej\supit{b}
\skiplinehalf
\supit{a}Steward Observatory, University of Arizona, 933 N. Cherry Avenue, 85721 Tucson, USA\\
\supit{b}D\'epartement d'Astrophysique, G\'eophysique et Oc\'eanographie, Universit\'e de Li\`ege, 17 All\'ee du Six Ao\^ut, B-4000 Sart Tilman, Belgium\\
\supit{c}Jet Propulsion Laboratory, California Institute of Technology 4800 Oak Grove Drive, Pasadena CA 91109-8099, USA\\
\supit{d}European Southern Observatory, Alonso de C\'{o}rdova 3107, Vitacura, Santiago, Chile\\
\supit{e}\AA{}ngstr\"{o}m Laboratory, Uppsala University, L\"{a}gerhyddsv\"{a}gen 1, SE-751 21 Uppsala, Sweden\\
\supit{f}Leiden Observatory, Niels Bohrweg 2, 2300 RA, Leiden, The Netherlands\\
\supit{g}Space Telescope Science Institute, 3700 San Martin Drive, Baltimore, MD 21218, USA\\
\supit{h}University of Virginia, Department of Astronomy, 530 McCormick Road, Charlottesville, VA 22904-4325, USA\\
}

\authorinfo{Send email correspondence to D.D. at ddefrere@email.arizona.edu.}
\begin{document} 
\maketitle 

%%%%%%%%%%%%%%%%%%%%%%%%%%%%%%%%%%%%%%%%%%%%%%%%%%%%%%%%%%%%% 
\begin{abstract}
We present the first observations obtained with the L'-band AGPM vortex coronagraph recently installed on LBTI/LMIRCam. The AGPM (Annular Groove Phase Mask) is a vector vortex coronagraph made from diamond subwavelength gratings. It is designed to improve the sensitivity and dynamic range of high-resolution imaging at very small inner working angles, down to 0.09 arcseconds in the case of LBTI/LMIRCam in the L' band. During the first hours on sky, we observed the young A5V star HR\,8799 with the goal to demonstrate the AGPM performance and assess its relevance for the ongoing LBTI planet survey (LEECH). Preliminary analyses of the data reveal the four known planets clearly at high SNR and provide unprecedented sensitivity limits in the inner planetary system (down to the diffraction limit of 0.09 arcseconds).
\end{abstract}
%>>>> Include a list of keywords after the abstract 

%%%%%%%%%%%%%%%%%%%%%%%%%%%%%%%%%%%%%%%%%%%%%%%%%%%%%%%%%%%%%
\section{INTRODUCTION}
\label{sec:intro}  % \label{} allows reference to this section

Recent advances in ground-based and space-based high-contrast direct imaging techniques have now produced the first direct images of extrasolar planets\cite{Marois:2008,Kalas:2008,Lagrange:2010,Kuzuhara:2013,Rameau:2013,Bailey:2014b}. Typically these are many-Jupiter-mass planets at moderate ($\sim$10 mag) to high ($\sim$15 mag) contrasts on relatively wide orbits ($>$0\farcs5 from the star). Direct imaging provides accurate measurements of the orbit, mass and composition of these planets, greatly improving our understanding of how planets form and evolve. One of the most remarkable systems discovered so far is HR\,8799, a young dusty planet-bearing A5V star located at 39\,pc\cite{Marois:2008,Marois:2010}. Not only it is one of the few planetary systems resolved by direct imaging, it is also the only one with multiple directly imaged planets. The multiple planet architecture around HR\,8799 provides a unique opportunity to study the physical and dynamical properties of a multi-planet system strikingly akin to a scaled up version of our outer Solar system in terms of mass ratios and planet/belts distances. Indeed, HR\,8799's four super-Jupiter planets are located at 15, 24, 38, and 68\,AU and are well nested within a double debris disk belt system. The inner belt, analogous to our asteroid belt is thought to be ranging from 6 to 15\,AU, while the outer one starts around 90\,AU, and spans hundreds of AUs\cite{Su:2009} which is reminiscent of our Kuiper belt and Oort cloud.  

Since the discovery of a fourth planet in 2010, the community of planet hunters has speculated about the presence of a fifth planet located closer in and shaping the inner debris disk. Several attempts at imaging the inner regions of HR\,8799 have already been made. The most successful so far are the non-redundant masking (NRM) Keck observations\cite{Hinkley:2011}, and the recent LBT observations by our team\cite{Skemer:2012,Skemer:2014}. The LBT data are the most sensitive so far and rule out planets as massive as the inner three ($e$,$d$,$c$) down to 0\farcs{235} (i.e., 9\,AU). This distance corresponds to the position of the mean motion resonance 2:1 with $e$, which is one of the possible stable configuration. In order to improve the detection limits at short angular separations, we installed an L'-band AGPM vortex coronagraph on LBTI/LMIRCam (see Section~\ref{sec:agpm}) and obtained observations of HR\,8799 on October 17th 2013 (see Section~\ref{sec:observations}). The vortex coronagraph is among the most promising solutions in that context, as it enables imaging down to the diffraction limit of the telescope (90\,mas at L' band or  $\sim$3.5\,AU at the distance of HR\,8799) and can be made achromatic over large bandwidths\cite{Mawet:2011,Mawet:2012}. We describe in this paper the results obtained (Section~\ref{sec:perfo}) and the near-term prospects (Section~\ref{sec:prospects}).

\begin{figure}[!t]
	\begin{center}
		\includegraphics[height=5.5 cm]{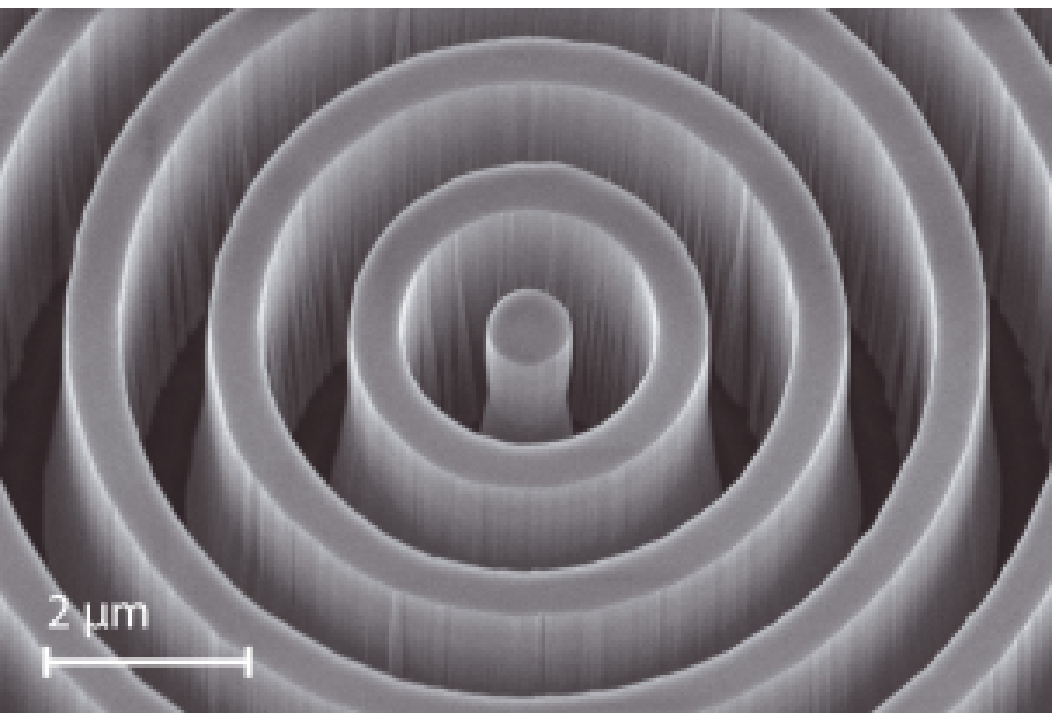}
	          \includegraphics[height=5.502 cm]{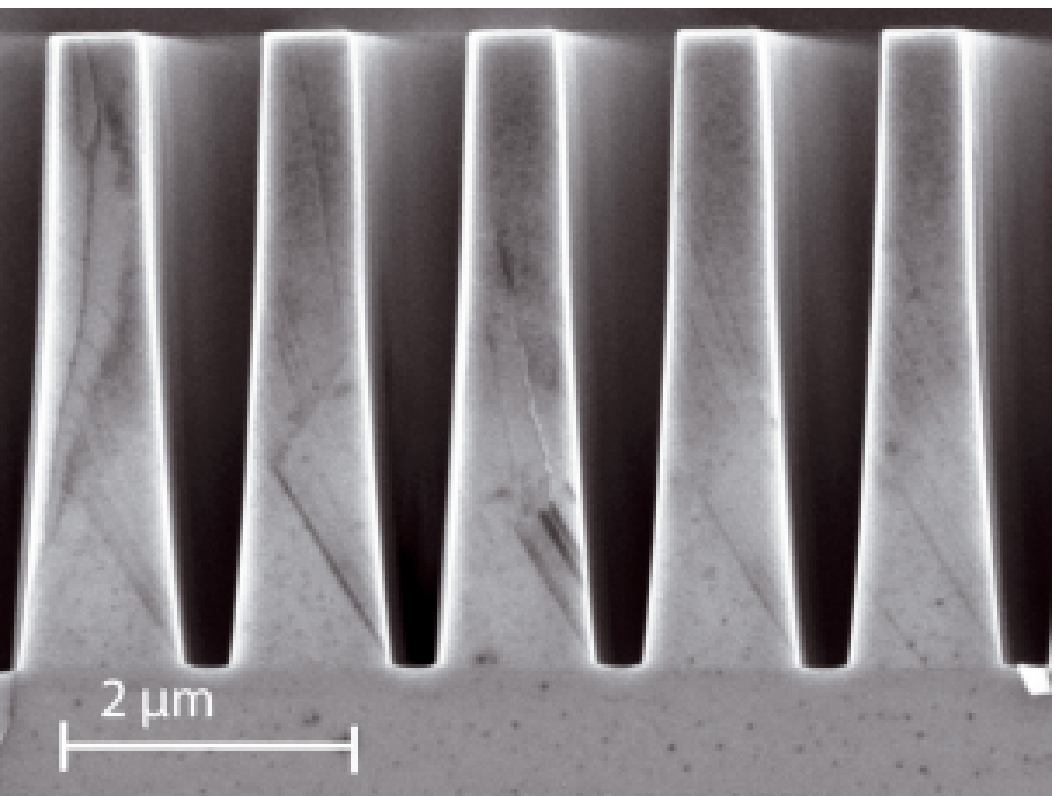}
		\caption{Left, scanning electron microscope (SEM) image of the center of an annular groove phase mask (AGPM). Right, SEM picture of the cleaved subwavelength grating, from which the geometric parameters of the AGPM-L4 profile are deduced: line width of 0.58\,$\mu$m and depth of 4.7\,$\mu$m. Both images are taken from Delacroix et al.~(2013)\cite{Delacroix:2013}.}\label{fig:grating}
	\end{center}
\end{figure}

\section{The AGPM Vector Vortex Coronagraph on LBTI/LMIRCam}\label{sec:agpm}

The Large Binocular Telescope\cite{Hill:2014,Veillet:2014} is a two 8.4-m aperture optical instrument installed on Mount Graham in southeastern Arizona (at an elevation of 3192 meters) and operated by an international collaboration between institutions in the United States, Italy, and Germany. Both telescopes are equipped with a deformable secondary mirror, which can be held in a fixed position for seeing limited instruments, or driven with the LBT's adaptive optics system to correct atmospheric turbulence at 1 kHz\cite{Bailey:2014,Christou:2014}. Each deformable mirror uses 672 actuators that routinely correct 400 modes and provide Strehl ratios exceeding 80\%, 95\%, and 99\% respectively at 1.6\,$\mu$m, 3.8\,$\mu$m, and 10\,$\mu$m\cite{Esposito:2010,Esposito:2012}.

The Large Binocular Telescope Interferometer (LBTI) is a NASA-funded nulling and imaging instrument designed to coherently combine the beams from the two primary mirrors of the LBT for high-sensitivity, high-contrast, and high-resolution infrared imaging (1.5-13\,$\mu$m). It is developed and operated by the University of Arizona and based on the heritage of the Bracewell Infrared Nulling Cryostat (BLINC) on the MMT\cite{Hinz:2000}. It is designed to be a versatile instrument that can image the two telescope beams separately, overlap the two beams incoherently, overlap the beams coherently for wide-field (Fizeau) interferometric imaging\cite{Leisenring:2014}, or overlap the beams and pupils for nulling interferometry\cite{Defrere:2014c}. It is equipped with two scientific cameras: LMIRCam\cite{Wilson:2008} (the L and M Infrared Camera) and NOMIC\cite{Hoffmann:2014} (Nulling Optimized Mid-Infrared Camera) covering respectively the 3-5\,$\mu$m and 8-13\,$\mu$m wavelength ranges. In addition, the near-infrared light (H and K bands) can be sent to a fringe tracker for high-angular resolution interferometric observations\cite{Defrere:2014}. The overall LBTI system architecture and performance are presented elsewhere in these proceedings\cite{Hinz:2014}.

The Annular Groove Phase Mask (AGPM) is an optical vortex made from diamond subwavelength gratings (see Figure~\ref{fig:grating}). It is designed to redirect the on-axis incident starlight outside the pupil where the light is blocked by a downstream Lyot stop. The idea of using an optical vortex as a focal-plane coronagraph was first proposed in 2005\cite{Mawet:2005a} and was historically driven from the need to have an achromatic phase mask with a continuous phase shift (no quadrant transition as in the case of the Four Quadrant Phase Mask\cite{Mawet:2003,Mawet:2005b}). Using subwavelength gratings made it possible to produce a continuous circular phase ramp around the optical axis\cite{Mawet:2005a} and the concept reached a sufficient readiness level for telescope implementation after eight years of intense technological development\cite{Absil:2014,Delacroix:2012,Delacroix:2013}. The AGPM coronagraph has many advantages over classical Lyot coronagraphs or phase/amplitude apodizers: small inner working angle, clear 360$^{\circ}$ off-axis field of view/discovery space, outer working angle set only by the instrument and/or mechanical/optical constraints, achromatic over the entire working waveband (here L' band), high throughput (here $\sim$88\%), and optical/operational simplicity. In January 2013, we installed on LMIRCam the fourth AGPM in a series of four realizations (AGPM-L4), the third one being installed on VLT/NACO\cite{Mawet:2013,Absil:2013c}. Its theoretical raw null depth limited by its intrinsic chromaticity was measured on a laboratory testbench to be around 5 $\times$ 10$^{-3}$ (corresponding to a raw contrast of 2.5 $\times$ 10$^{-5}$ at $2\lambda/D$), which is more than needed for on-sky operations where the limit is set by the residual wavefront aberrations. More information about the AGPM and the VORTEX project can be found elsewhere in these proceedings\cite{Absil:2014,Delacroix:2014,Carlomagno14,Forsberg14,Jolivet14}.

 \begin{figure}[!t]
\centering
\includegraphics[width=12.0 cm]{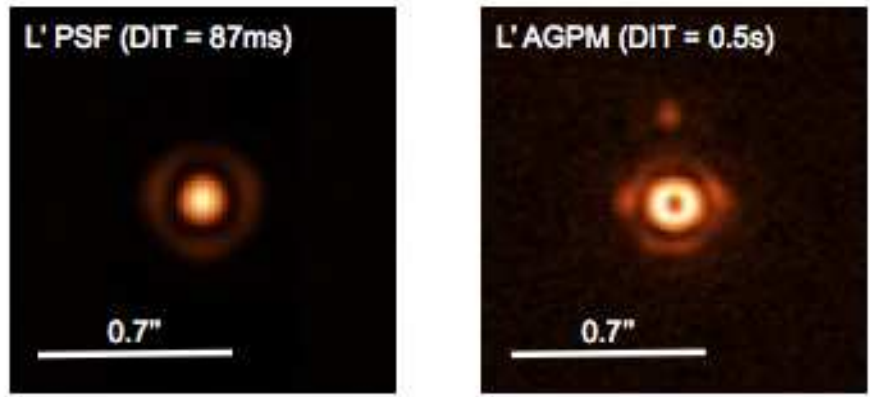}
\caption{L'-band LBTI/LMIRCam PSF in direct imaging (left) and L'-band LBTI/LMIRCam coronagraphic image with the star centered on the AGPM (right, data from October 17th, 2013). For this first try, the rejection ratio of the stellar flux was approximately 35 (far from optimal). Based on our experience with other telescopes, we expect to reach at least 100 next time by doing a better optimization of the telescope focus. The bright spot above the coronagraphic PSF and visible in the right image is due to a secondary reflection inside LMIRCam. The color scale of the right image has been modified to better show the coronagraphic PSF. } \label{fig0}
 \end{figure}
 
\section{Observations and data reduction}\label{sec:observations}

We observed HR\,8799 on October 17th 2013 during the first commissioning run of the AGPM. The observations were obtained around transit  during approximately 3.5 hours. We used only the left-side telescope and blocked the right side which was used by another non-LBTI camera. The seeing was fair during the first 30 minutes (1.2''-1.4'') and good for the remaining of the observations (0.9''-1.0''). The AO loop was locked with 200 modes first and with 400 modes after 30 minutes. The observing sequence was composed of 1000 frames of 0.5s each followed by a sequence of 300 frames of nearby empty sky region with the same integration time. The star position was centered at the beginning of each sequence to achieve the maximal extinction. The non-coronagraphic PSF through the Lyot stop and AGPM was regularly calibrated during the observing sequence by placing the star a few arcsecs away  from the vortex center.  

In order to remove the relative offsets between camera readout channels due to time-dependent voltage drifts, the raw images were first bias-corrected using reference pixels located at the top and bottom of each frame. Similar integration dark frames, taken at the end of the night, were then subtracted and the resulting images flattened. Background subtraction was then performed using the median-combination of corresponding background frames. Using a master bad pixel mask generated with sky data taken at the end of night, the bad pixels were subsequently fixed in each frame using the median of adjacent pixels. The images were then centered by bidimensional Moffat fit of LMIRcam's secondary reflection (see Figure~\ref{fig0}), making sure that the star is placed at the same exact position in all individual frames. Bad frames affected by strong AO loop openings were removed using a 3-$\sigma$ clip of the fitted Moffat slope profile. Finally, a new image cube was created by binning each image, 2$\times$2, and averaging 20 successive frames (i.e., 10 s of effective integration time). 

\begin{figure}[!t]
\centering
\includegraphics[width=15.0 cm]{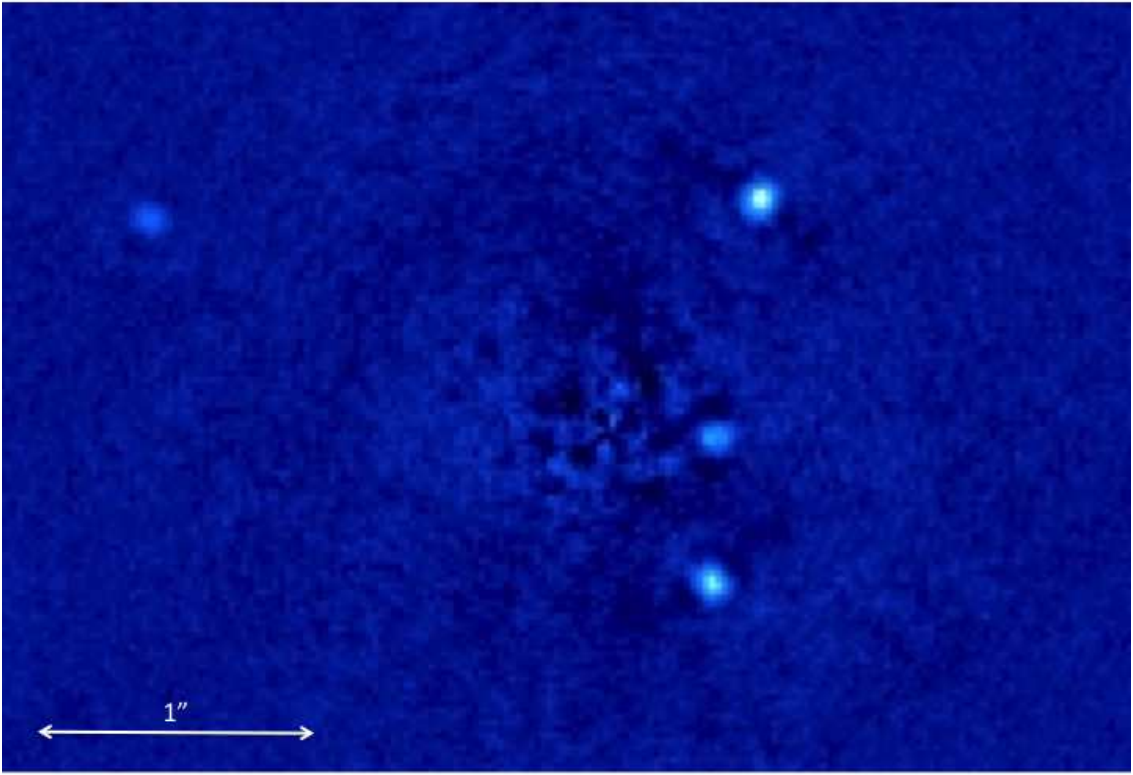}
\caption{First L'-band coronagraphic image (linear scale) obtained with LMIRCam/AGPM on 2013 October 17th. The four known planets around HR 8799 are all detected clearly at high SNR (see contrast curve in Figure~\ref{fig2}). The central star has been subtracted by the AGPM coronagraph and image processing. The black spots on either sides of the planet are artefacts related to the rotation of the planet around the optical axis in the image sequence.} \label{fig1}
 \end{figure}

In order to produce a final image of the HR\,8799 system, we used and compared three independent reduction pipelines. Taking advantage of angular differential imaging (ADI), they are all based on the KLIP algorithm\cite{Soummer:2012} in which the ADI image sequence is used as a PSF library, to which a principal component analysis (PCA) treatment is applied. We also tried the so-called ``smart" version of the PCA in which the image library is built only from images where the off-axis companion has rotated by at least $\lambda$/D with respect to the image under consideration\cite{Absil:2013c}. All approaches led to similar results. The final image obtained using the pipeline developed by the Vortex team\cite{Mawet:2013,Absil:2013c} is shown in Figure~\ref{fig1} and reveals the four known planets around HR 8799 clearly at high SNR.
 
\begin{figure}[!ht]
\centering
\includegraphics[width=15.5 cm]{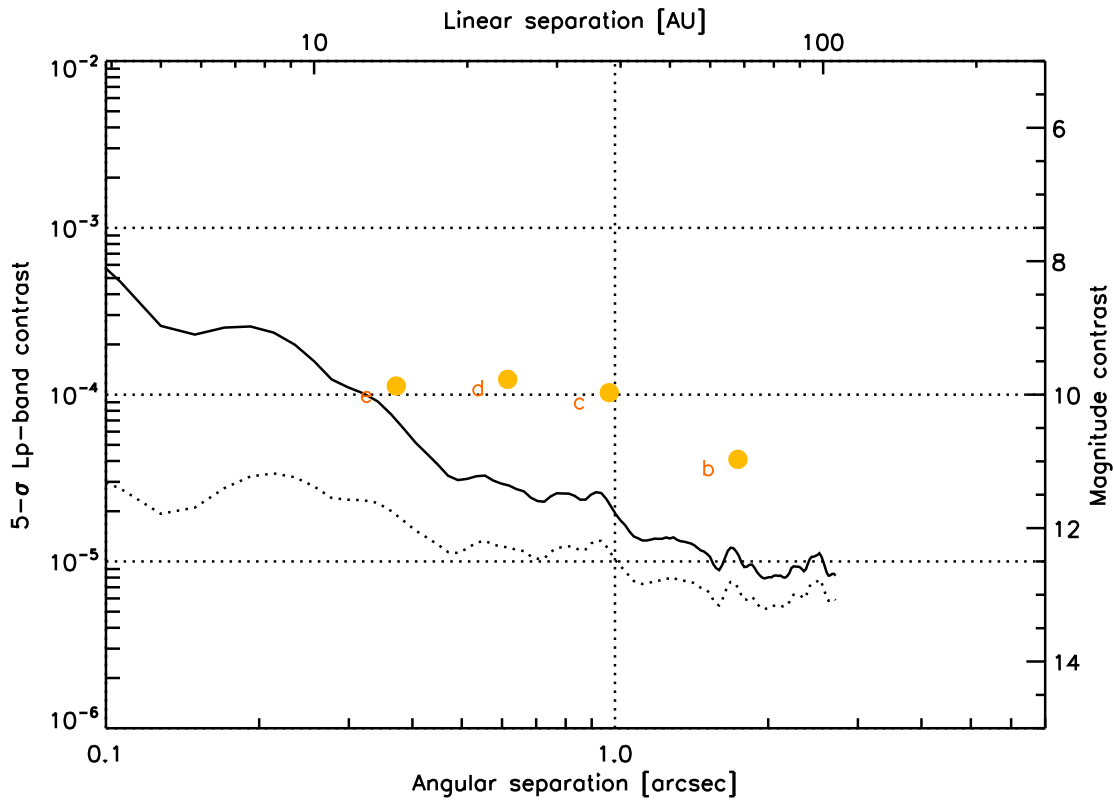}
\caption{Current 5-$\sigma$ detectability of LMIRCam/AGPM in terms of contrast  for point-like companions around HR\,8799 (solid line). The dotted line represents the raw contrast curve (without considering self-subtraction). The four-known planets are shown for  comparison (magnitude from the discovery papers\cite{Marois:2008, Marois:2010}).} \label{fig2}
\end{figure}

\section{Current performance}\label{sec:perfo}

The performance of high-contrast direct imaging instruments is generally estimated in terms of achievable contrast for point-like companions as a function of the angular separation from the star. We estimate here the noise level as the standard deviation of the pixel intensity in concentric annuli, or equivalently as the azimuthal median of the noise computed locally in small square boxes. The amount of self-subtraction, inherent to ADI observations, was estimated by introducing fake companions directly in the data cube, separated by a few $\lambda$/D from each other and placed on three radial branches separated by 120$^{\circ}$  in azimuth. The fake companions were injected at 20$\sigma$ above the noise computed after a first pass of the PCA algorithm on the cube without fake companions. By measuring the photometry of the fake companions in the final reduced image after a second pass of the PCA algorithm and comparing it to their input flux, we inferred the attenuation of the PCA algorithm in ADI mode. The final contrast curve, computed as five times the standard deviation of the pixel intensities in $\lambda$/D-wide annuli after applying a median filter on a $\lambda$/D-wide moving box, is displayed in Figure~\ref{fig2} before and after taking into account companion self-subtraction. This contrast curve is very similar to the one obtained in direct imaging mode by LMIRCam (no coronagraph) under similar observing conditions\cite{Skemer:2014}. The contrast achieved inside the radius of HR\,8799e will be validated by looking at a well-known binary during our next observations. As mentioned before, the rejection\footnote{The rejection ratio is defined here as the ratio between the maximum intensity in the direct image and the maximum intensity in the coronagraphic image.} ratio of the stellar flux for these first AGPM observations was only approximately 35 which is far from optimal. Based on our experience with other telescopes, we expect to reach a rejection factor of 100 next time by doing a better adjustment of focus in the AGPM plane.

\section{Summary and future work}\label{sec:prospects}

In this paper, we have presented the first observations obtained with the AGPM vortex coronagraph recently installed on LBTI/LMIRCam. The goal of these observations was to demonstrate the exquisite inner working angle of the AGPM  and assess its relevance for the ongoing LBTI planet survey (LEECH, LBTI Exozodi Exoplanet Common Hunt \cite{Skemer:2014}). The results of our first attempt show the four know planets around HR\,8799 clearly at high SNR despite a coronagraphic starlight rejection ratio of only $\sim$35. Future work and observations will be focused on improving the rejection ratio (we expect to reach a rejection ratio of at least 100 based on our experience with other telescopes). In order to improve the performance, we plan to implement an automatic centering loop based on the science frames and a better algorithm to adjust the focus in the AGPM plane. In addition, we are currently investigating the possibility of installing multiple AGPMs inside LMIRCam to take advantage of both telescopes and get the necessary redundancy which is crucial to distinguish planets from false detections at small angular separations. In parallel, the Vortex team is currently developing a more advanced version of PCA-based processing and designing subwavelength gratings of higher topological charge to improve the AGPM rejection\cite{Delacroix:2014}. Combined with LBT's state-of-the-art high-performance adaptive optics system and taking full advantage of LBTI 's extremely low thermal background, we expect LMIRCam/AGPM to open a completely new parameter space for high-contrast imaging at L', a wavelength optimized for detecting cold, low-luminosity exoplanets.

%%%%%%%%%%%%%%%%%%%%%%%%%%%%%%%%%%%%%%%%%%%%%%%%%%%%%%%%%%%%%
\acknowledgments     %>>>> equivalent to \section*{ACKNOWLEDGMENTS}       
LBTI is funded by a NASA grant in support of the Exoplanet Exploration Program (NSF 0705296). The LBT is an international collaboration among institutions in the United States, Italy and Germany. LBT Corporation partners are: The University of Arizona on behalf of the Arizona university system; Istituto Nazionale di Astrofisica, Italy; LBT Beteiligungsgesellschaft, Germany, representing the Max-Planck Society, the Astrophysical Institute Potsdam, and Heidelberg University; The Ohio State University, and The Research Corporation, on behalf of The University of Notre Dame, University of Minnesota and University of Virginia. The research leading to these results has received funding from the European Research Council under the European Union's Seventh Framework Programme (ERC Grant Agreement n.337569) and from the French Community of Belgium through an ARC grant for Concerted Research Actions. O.A. is a Research Associate of the F.R.S.-FNRS (Belgium).

%%%%%%%%%%%%%%%%%%%%%%%%%%%%%%%%%%%%%%%%%%%%%%%%%%%%%%%%%%%%%
%%%%% References %%%%%

\bibliographystyle{spiebib}   %>>>> makes bibtex use spiebib.bst

%\subsection{Internal source}

%The LBTI has two artificial point sources that can be used for phase sensing tests.  The first is a small NiChrome wire within the NIC to test the nulling interferometer and phase sensor, as shown in Figure 2. The second is a superluminescent diode source at 1.55\,$\mu$m that %can be used to do an end-to-end test of the FPC and phasecam phase sensor. 

%Precise pathlength adjustments can be put in with two PZT devices. Each have a throw of 11$\mu$m, or 22$\mu$m in OPD.

\end{document}